\documentclass{ws-ijmpe}
\usepackage[super,compress]{cite}
\begin{document}


\catchline{}{}{}{}{}

\title{APPROACH TO EQUILIBRIUM IN THE CALDEIRA-LEGGETT MODEL }


\author{\footnotesize VENKITESH AYYAR and BERNDT M\"ULLER}

\address{Department of Physics, Duke University, Durham, NC 27708-0305, USA}

\maketitle

\begin{history}
\end{history}

\begin{abstract}
The Caldeira-Leggett model describes a microscopic quantum system, represented by a harmonic oscillator, in interaction with a heat bath, represented by a large number of harmonic oscillators with a range of frequencies. We consider the case when the system oscillator starts out in the ground state and then thermalizes due to interactions with the heat bath, which is at temperature $\theta$. We calculate the position autocorrelation function $\langle x(t')x(t) \rangle$ of the system oscillator at two different times and study its behaviour in the small and large time limits. Our results show that the system oscillator thermalizes as expected. We also confirm by explicit calculation that the position autocorrelation function exhibits periodicity for imaginary values of the time difference $t'-t = i\tau$ at late (real) times $t$.
\end{abstract}




\section{ Introduction }

The Langevin equation provides a good description of many dissipative classical systems. One approach to finding the quantum analogue of the Langevin equation is to quantize the system interacting with an environment as a whole, and then to eliminate the environmental degrees of freedom. The interaction between the system and the environment gives rise to dissipation. The Caldeira-Leggett (C-L) model \cite{1} follows this approach, modeling the environment as a large collection of thermal harmonic oscillators interacting linearly with the system, which is also modeled as a harmonic oscillator. The interaction between the system and the environment is assumed to be linear in both coordinates. The parameters of the quantum model are set after comparison with the classical equation in the appropriate limit. 

Because the C-L model is simple enough to be analytically solved, it is possible to study the approach of the system oscillator to thermal equilibrium in detail and to explore how the properties of thermal equilibrium are attained at late times. For example, we will ask how the periodicity of Green functions in imaginary time of the thermal system emerges from the real-time evolution of the system interacting with the environment. We show that the periodicity only holds on the compactified interval $0 \leq \tau \leq 1/\theta$, where $\theta$ is the temperature of the environment. For convenience we choose natural units, in which the Boltzmann constant and Planck's constant are set to one: $\hbar = k_B = 1$.

\section{The C-L Model}

Consider a particle of mass $M$ in a potential $v(x)$ and in contact with a reservoir consisting of large number of harmonic oscillators each with mass $m$ and frequencies $ \omega_k $.  Let $x$ be the position of the particle and $ R_k $ be those of the reservoir. The interaction between the particle and the environment is assumed to be linear in the coordinates. The Hamiltonian of the model is:
\begin{eqnarray}
H &=& H_{\rm sys} + H_{res} + H_{int} \nonumber \\
 &=& \frac{p^2}{2 M} + v(x) + x \sum_k C_k R_k + \sum_k \frac{p_k^2}{2 m} + \sum_k \frac{1}{2} m \omega_k^2 R_k^2 
\end{eqnarray}
where $v(x) = \frac{1}{2} m \omega^2 x^2$ is the potential. The particle trapped in the potential $v(x)$ is called the ``system''. The reservoir is assumed to be in thermal equilibrium at a temperature $ \theta $ and acts as a heat bath.

Using the notation $ \rho(x,y,t) $ for the $ (x,y)-$element of the reduced density matrix of the system, the reduced density operator at a time $ t $ is given by 
\begin{equation} 
\rho(x,y,t) = \int dx' dy' J(x,y,t,x',y',0) \rho(x',y',0) .
\end{equation}  
The propagator $ J $ of the density operator can be obtained using the Feynman-Vernon influence functional method (see eq.~(3.38) of Ref. \refcite{1}):
\begin{equation}
J(x,y,t;x',y',0) =
\int { D x D y}  \exp \left( i S_0[x,y] - S_1[x,y] \right) 
\end{equation}
with
\begin{eqnarray}
S_0[x,y] &=& S_R[x] - S_R[y] -M \gamma \int_0^t (x \dot{x}) -y \dot{y} + x \dot{y} - y \dot{x} ) d \tau
\\
S_R &=& \int_0^t \left[ \frac{1}{2} M \dot{x}^2 -v(x) + \frac{1}{2} M (\Delta \omega)^2 x^2 \right] d \tau
\\
S_1[x,y]
&=&  \frac{2 M \gamma }{\pi} \int_0^ \Omega d\omega'\, \omega' \coth \left( \frac{ \omega' }{2 \theta } \right) 
\nonumber \\  
& & \times 
\int_0^t \int_0^{\tau} [x(\tau)-y(\tau)] \cos[\omega' (\tau - s)] [x(s)-y(s)] d\tau ds  ,
\end{eqnarray}
where $\gamma $ is the relaxation constant of the system, and $ \Omega $ is the upper frequency cutoff for the reservoir.  The natural frequency $ \omega $ of the oscillator is modified by the interaction. The renormalized frequency of the system oscillator is
\begin{equation}
{\omega_R}^2 = {\omega}^2 - {(\Delta \omega)}^2 
\end{equation}
with
\begin{equation}
{(\Delta \omega)}^2 = \frac{4 \gamma \Omega}{\pi} .
\end{equation}
In addition, it is useful to introduce the reduced frequency of oscillation of the system in the presence of damping:
\begin{equation}
w = {\omega_R}^2 - \gamma^2 .
\end{equation}

In general, the relaxation constant $ \gamma $ depends on the couplings $ C_k $ in a complicated manner. If the environment consists of $ N $ oscillators and every $ C_k  = C $, 
using equations (3.34) and ( 3.36) of \cite{1}  , we get 
\begin{eqnarray}
 \gamma = \frac{ 3 N C^2 \pi }{4 m M \Omega^3}
\end{eqnarray}

\section{Approach to thermal equilibrium}

\subsection{System behaviour at the large times}

For an undamped harmonic oscillator of frequency $ \omega $, the propagator of the density operator is given by (see eq.~(6.26) of  Ref. \refcite{1}):
\begin{eqnarray}
 J &=& F^2 (t) \exp \left[ i \left(K_2(t) X_f \xi_f + K_1(t) X_i \xi_i - L(t) X_i \xi_f - N(t) X_f \xi_i \right) \right] \\
&& \times \exp \left[ - \left( A(t) {\xi_f}^2 + B(t) \xi_f \xi_i + C(t) {\xi_i}^2) \right) \right] ,
\label{eq:J}
\end{eqnarray}
where 
\begin{equation} 
F^2 (t) = \lim_{N \to \infty} \frac{4 \tilde{C}}{M t} (2 \pi)^N \prod_{i=1}^N \frac{1}{{\omega_i}^2 -{\omega_R}^2} .
\end{equation}
$  A, B, C, K, L, N $ are functions of time and $ K_{1 / 2} = K \pm { M \gamma / 2} $. 
Here, $ J $ has been written in terms of the sum and difference coordinates  $ X = x+y $ and $ \xi = x - y $ respectively, instead of $ x $ and $ y $.

The initial density matrix for a simple harmonic oscillator in the ground state is 
\begin{equation} 
\rho(x_i,y_i,0) = {1 \over {\sqrt \pi x_0}} \exp { \left[ -(x_i ^2 + y_i ^2 ) \over { 2 x_0^2 } \right] }
 \end{equation}
where $x_0 = (M \omega)^{-1/2} $.  In terms of the sum and difference coordinates $ X $ and $ \xi $,  
\begin{equation} 
\rho(X_i,\xi_i,0) = {1 \over {\sqrt \pi x_0}} \exp{ \left[ -(X_i ^2 + \xi_i ^2 ) \over {4 x_0 ^2 } \right] }
 \end{equation}
The density matrix at time $ t $ is given by
\begin{eqnarray} \nonumber
\rho(X_f , \xi_f , t) &=& {F^2 \over \sqrt{\pi} x_0 } \exp(i K_2 X_f \xi_f - A \xi_f ^2 ) \\ 
\nonumber 
&&\int{dX_i d\xi_i} \exp \left[ - \xi_i ^2 \left(C + {M \omega \over {4}}\right) 
- X_i ^2 \left({M \omega \over 4}\right)+ X_i \xi_i (i K_1)\right] \\ 
&& \times \exp \left[ X_i (-i L \xi_f)+ \xi_i (-iN X_f - B \xi_f ) \right] \phantom{\frac{1}{2}}  
\end{eqnarray}
This Gaussian integral can be solved by applying the general formula:
\begin{eqnarray}
 \int \exp{\left[ -\frac{{\bf x}\cdot{\bf A}\cdot{\bf x}}{2}+{\bf J}\cdot{\bf x}\right] } d^n x 
 = \sqrt{\frac{(2 \pi )^n}{\det{\bf A}}} \exp{\left\{ {{\bf J}\cdot{\bf A}^{-1}\cdot{\bf J}\over 2 }\right\}}  ,
\end{eqnarray}
where ${\bf x}$ and ${\bf J}$ are vectors and ${\bf A}$ is a symmetric $(n\times n)$ matrix.
For $ n= 2 $, we obtain the density matrix in the final state as 
\begin{eqnarray}
  \rho(X_f,\xi_f,t) &=& F^2 \sqrt{16 \pi M \omega / z} \exp\left[- ( a {\xi_f}^2 + b {X_f}^2 + i c X_f \xi_f ) \right] 
  \label{eqn:1} 
\end{eqnarray}
where
\begin{eqnarray}
 z &=& 4 {K_1}^2 +M \omega (4 C + M \omega )\label{eqn:z} \\
 a &=& A + \frac{4 C L^2 - B^2 M \omega + L^2 M \omega +4 B K_1 L}{z} \label{eqn:a}  \\
 b &=&  M N^2 \omega \over z  \label{eqn:b} \\
 c &=& -K_2 + \frac{4 K_1 L N - 2 B M N \omega}{z} \label{eqn:c} 
\end{eqnarray}
Thus the density matrix element $ \rho(x,y,t) $  is a Gaussian.

\begin{figure}[htb]
\centering
\includegraphics[width=0.48\linewidth]{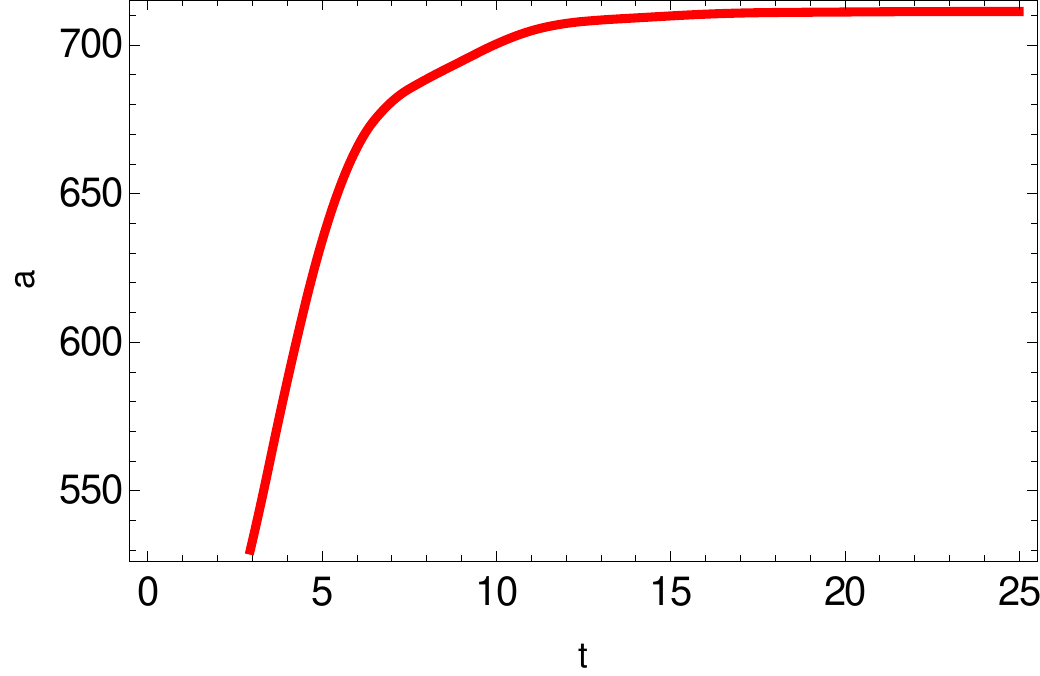}
\hfill
\includegraphics[width=0.48\linewidth]{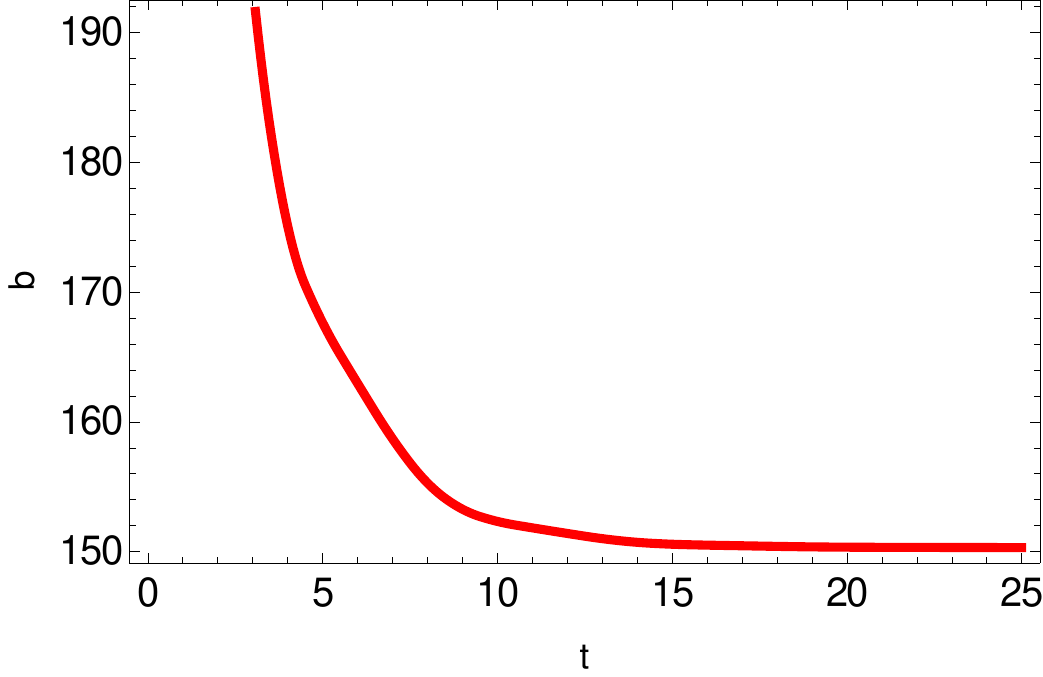}
\caption{The coefficients  $a(t)$ and $b(t)$. Both coefficients attain a constant value at late times.}
\label{fig:f1f2}
\end{figure}

\begin{figure}[htb]
\centering
\includegraphics[width=0.75\linewidth]{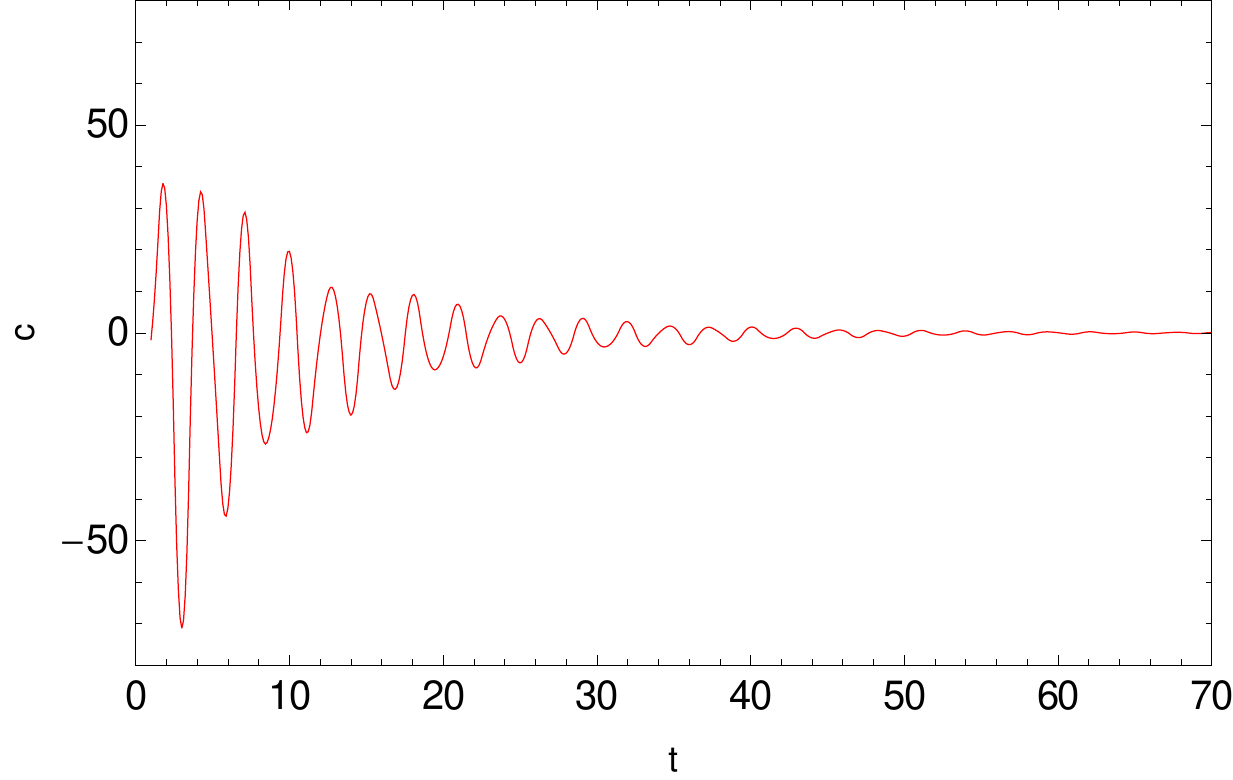}
\caption{Oscillatory behaviour of the coefficient $c(t)$, which approaches zero at late times.}
\label{fig:f3}
\end{figure}

Plots of the variation of $a$, $b$, and $c$  with time are shown in Figs.~\ref{fig:f1f2}, \ref{fig:f3} for the parameters
\begin{equation} 
\theta = 10~ \mbox{MeV} , \gamma = 0.2~ \mbox{MeV}, \Omega = 100~ \mbox{MeV} , M = 130~ \mbox{MeV} .
\end{equation}
We observe that $a$, $b$ and $c$ stabilize after some time as what one would expect for a system that thermalizes.
We define the relaxation time to be $ t_{\rm rel} = (2 \gamma)^{-1/2} $, which for our choice of parameters is $  t_{\rm rel} = 2.5~ (\mbox{MeV})^{-1} $.
The system is very close to thermal equilibrium after about 8 relaxation times, i.e.\ after about $ 20~ (\mbox{MeV})^{-1} $.


Using the normalization condition for the reduced density matrix $ \rho(t) $ , we can express $ F $ in terms of  $ N $ as follows.  Since the density matrix at $ t = 0 $ is normalized to one, unitarity dictates that it continues to remain normalized at any time $ t $:
\begin{equation}
{\rm Tr}[\rho(t)] =\int{ dx\, \rho(x,x,t)}= 1 
\label{eqn:2}
\end{equation} 
From (\ref{eqn:1}) we get:
\begin{equation}
\rho_{f} ( x,x,t) = \rho_f (X,0,t) 
= F^2 (t) \sqrt{16 \pi M \omega / z} \exp \left( -\frac{M N^2 \omega}{z} \right) ;
\end{equation}
upon substitution into (\ref{eqn:2}) we obtain 
\begin{equation}
\int dx\,  F^2 \sqrt{16 \pi M \omega / z } \exp\left(- {4 M N^2 \omega \over z} x^2 \right) 
= \frac{ 2\pi F^2 }{ |N| } = 1 ,
\end{equation}
resulting in the normalization condition
\begin{equation}
F^2 (t) = \frac{|N(t)|}{2 \pi } .
\end{equation}

\subsection{Harmonic oscillator at a temperature $ \theta $ }

Consider now the system oscillator in thermal equilibrium with temperature $\theta$. Its density matrix is given by:\cite{3}
\begin{eqnarray}
\rho(x,y) &=& \frac{M w}{2 \pi \sinh( w / \theta)} \nonumber \\
 &&\exp{\left[ - \frac{M w}{2 \sinh( w / \theta)} 
 \left( (x^2 + y^2) \cosh( w / \theta)  -2 x y   \right) \right]} ,
\end{eqnarray}
In terms of the sum and difference coordinates  $ X $ and $ \xi $ the density matrix can be written as 
\begin{equation}
 \rho(X,\xi) =  \frac{M w}{2 \pi \sinh(w / \theta)}  \exp{\left( - a_{\rm th} \xi^2 - b_{\rm th} X^2 \right)} 
\end{equation}
where
\begin{eqnarray}\nonumber
 a_{\rm th} &=& \frac{M w}{4 \sinh(w / \theta)}( \cosh(w / \theta) + 1 ), \\
 b_{\rm th} &=& \frac{M w}{4 \sinh(w / \theta)}( \cosh(w / \theta) - 1 ) 
 \label{eqn:11}
\end{eqnarray}
\\

\subsection{Comparison of the C-L model with a thermal system}

We now can compare the values of $ a $ and $ b $  for the thermalized system with those obtained for the model at late times ( $t = 8 t_{\rm rel} = 20~ (\mbox{MeV})^{-1} $) . Comparison of  $ a(t) $ and $ b(t) $ given by eqs. (\ref{eqn:a}, \ref{eqn:b}) with $a_{\rm th}$ and $ b_{\rm th} $ given by eq. (\ref{eqn:11}) confirms that the system oscillator in the C-L model thermalizes, indeed. Table 1 provides a comparison of the dynamically obtained coefficients with the thermal ones for various values of $ \gamma $. Table 2 gives a similar comparison for various values of temperature $ \theta $.

\begin{table}[htb]
\tbl{Comparison of the C-L model at $t=8t_{\rm rel}$ with a thermalized oscillator for various values of the damping rate $ \gamma$.}
{\begin{tabular}{@{}ccccc@{}} \toprule
$\gamma  $ & $ a $  &  $ a_{\rm th} $  & $ b $ & $ b_{\rm th} $  \\
(MeV) & $ {\mbox{(MeV)}}^2 $ & $ {\mbox{(MeV)}}^2 $  & $ {\mbox{(MeV)}}^2 $ & $ {\mbox{(MeV)}}^2 $ \\  \colrule
0.10\hphantom{00} & \hphantom{0}700.149 & \hphantom{0}703.285 & 151.659 &  150.188 \\
0.15\hphantom{00} & \hphantom{0}708.309 & \hphantom{0}703.285 & 150.452 &  150.188\\
0.20\hphantom{00} & \hphantom{0}711.141 & \hphantom{0}703.285 & 150.329 &  150.188\\
0.25\hphantom{00} & \hphantom{0}713.235 & \hphantom{0}703.285 & 150.353 &  150.188\\
0.30\hphantom{00} & \hphantom{0}715.222 & \hphantom{0}703.285 & 150.403 &  150.188\\
0.35\hphantom{00} & \hphantom{0}717.187 & \hphantom{0}703.285 & 150.464 &  150.188\\
 \botrule
\end{tabular}}
\begin{tabfootnote}
\end{tabfootnote}
\end{table}

\begin{table}[htb]
\tbl{Comparison of the C-L model at $t=8t_{\rm rel}$ with a thermalized oscillator for various values of the temperature $ \theta $.}
{\begin{tabular}{@{}rrrrr@{}} \toprule
$\theta$\hphantom{00} & $a$\hphantom{000}  &  $a_{\rm th}$\hphantom{00}  & $b$\hphantom{000} 
& $b_{\rm th}$\hphantom{00}  \\
(MeV) & $ {\mbox{(MeV)}}^2 $ & $ {\mbox{(MeV)}}^2 $  & $ {\mbox{(MeV)}}^2 $ & $ {\mbox{(MeV)}}^2 $ \\ \colrule
10\hphantom{00} &  711.14\hphantom{0}  & 703.29\hphantom{0} & 150.33 &  150.188 \\
15\hphantom{00} & 1015.76\hphantom{0} & 1010.85\hphantom{0} & 104.57 &  104.492\\
20\hphantom{00} & 1329.25\hphantom{0} & 1326.97\hphantom{0} &   79.655  &   79.599\\
25\hphantom{00} & 1646.94\hphantom{0} & 1646.61\hphantom{0} &   64.192  &   64.147\\
30\hphantom{00} & 1966.68\hphantom{0} & 1968.02\hphantom{0} &   53.708  &   53.670\\ \botrule
\end{tabular}}
\begin{tabfootnote}
\end{tabfootnote}
\end{table}

\section{Position correlation function}

\subsection{Two-time correlation function}

We now calculate the two point correlation function of the position operator for the system oscillator at different times $\langle x(t+\tilde{t}) x(t) \rangle$.  The time evolution of the density operator $\rho_s$ in the Schr\"odinger picture can be expressed in two different ways:
\begin{equation}
\rho_s(t) = {\rm Tr }_{\rm env} \left[ U(t) \rho_{\rm sys}(0) \rho_{\rm env}(0) U^{\dagger}(t) \right] , 
\label{eqn:r1}
\end{equation}
where $U(t) = \exp{\left[-i H t\right]}$ is the time evolution operator and 
\begin{equation}
\rho_{\rm sys}(t) = J(t) \rho_{\rm sys}(0) .
\label{eqn:r2}
\end{equation}
The density matrices of the system and the environment are assumed to factorize at $t=0$.

We can therefore write the $(x,y)$ element of the density matrix in the following two equivalent forms. First,
simplifying (\ref{eqn:r1}), we obtain
\begin{eqnarray}
\langle x|\rho_{\rm sys}(t)|y\rangle &=& {\rm Tr }_{\rm env} \left[ \int{ \langle x|U(t)|x'\rangle  \langle x'|\rho_{\rm sys}(0)|y'\rangle \rho_{\rm env}(0) \langle y'|U^{\dagger} (t) | y\rangle  dx' dy' }\right] \nonumber \\ 
&=& \int{dx' dy'  \langle x'|\rho_{\rm sys}(0)|y'\rangle {\rm Tr }_{\rm env} \left[ \langle x|U(t)|x'\rangle \rho_{\rm env}(0)  \langle y'|U^{\dagger} (t) | y\rangle  \right]  }
\end{eqnarray}
Alternatively, from eq. (\ref{eqn:r2}) we get
\begin{eqnarray}
\langle x|\rho_{\rm sys}(t)|y\rangle = \int { J(x,y,x',y',t,0)\langle x'|\rho_{\rm sys}(0)|y'\rangle  dx' dy' } 
\end{eqnarray}
Comparing these two forms, we obtain the following expression for the propagator $J$ of the density matrix in terms of the time evolution operator:
\begin{equation}
J(x,y,x',y',t,0) = {\rm Tr }_{\rm env} \left[ \langle x|U(t)|x'\rangle \rho_{\rm env}(0)  \langle y'|U^{\dagger}(t)|y\rangle \right] 
\label{eqn:rho1}
\end{equation}
We can use this relation to express the time evolution of the Heisenberg picture operator ${\bf x}(t)$ in terms of the propagator $J$. Since $ {\rm Tr}_{\rm env} \left[ \rho_{\rm env}(0) \right]= 1 $,
\begin{eqnarray}
\langle x|x(t)|y\rangle &=& \langle x| x(t)|y\rangle {\rm Tr }_{\rm env} \left[ \rho_{\rm env}(0) \right] \nonumber \\ 
&=& {\rm Tr }_{\rm env}\left[ \langle x| x(t) \rho_{\rm env}(0)|y \rangle \right] 
\label{eqn:r3}
\end{eqnarray}
Substituting $x(t) = U^{\dagger}(t) x(0) U(t)$  in Eqn \ref{eqn:r3}, we obtain :
\begin{eqnarray}
\langle x|x(t)|y\rangle  
&=& {\rm Tr }_{\rm env}\left[ \int{ dx' dy' \langle  x|U^{\dagger}(t)|x'\rangle  \langle x'|x(0)|y'\rangle \langle y'|U(t)|y\rangle \rho_{\rm env} (0)} \right]\nonumber \\
&=& {\rm Tr }_{\rm env}\left[  \int{ dx' dy'\langle  x|U^{\dagger}(t)|x'\rangle  x' \delta_{x',y'}\langle y'|U(t)|y\rangle  \rho_{\rm env} (0) } \right]\nonumber \\
&=& {\rm Tr }_{\rm env}\left[ \int {dx' x' \langle x|U^{\dagger}(t) | x'\rangle  \langle x'| U(t) | y\rangle   \rho_{\rm env} (0) }  \right]\nonumber \\ 
&=& {\rm Tr }_{\rm env}\left[ \int {dx' x' \langle x'| U(t) | y\rangle   \rho_{\rm env} (0) \langle x|U^{\dagger}(t) | x'\rangle  }  \right] 
\end{eqnarray}
In view of (\ref{eqn:rho1}) this implies:
\begin{equation}
\langle x|x(t)|y\rangle  = \int { dx' x' J(x',x',y,x,t,0)} \label{eqn:rho2}
\end{equation}

We are now ready to calculate the two point correlation function of the position of the system:
\begin{eqnarray}
 \langle x(t+\tilde{t}) x(t)\rangle  &=& {\rm Tr }\left[ \rho (0) x(t+\tilde{t}) x(t) \right] \nonumber \\
&=& {\rm Tr }_{\rm sys} \left[ {\rm Tr }_{\rm env} \left\{  \rho (0) U^{\dagger} (t+\tilde{t}) x(0) U(t+\tilde{t}) U^{\dagger}(t) x(0) U(t) \right\} \right] \nonumber\\ 
&=& {\rm Tr }_{\rm sys} \left[ {\rm Tr }_{\rm env} \left\{  U(t) \rho (0) U^{\dagger} (t) U^{\dagger} (\tilde{t}) x(0) U(\tilde{t}) x(0) \right\} \right] \nonumber \\ 
&=& {\rm Tr }_{\rm sys} \left[ {\rm Tr }_{\rm env} \left\{ \rho (t) x(\tilde{t}) x(0) \right\} \right] \nonumber\\ 
&=&  {\rm Tr }_{\rm env} \left[ \int{ dx dy \langle x|\rho (t)|y\rangle  \langle y|x(\tilde{t})|x\rangle  x} \right] \nonumber\\ 
&=& \left[ \int{ dx dy  \langle y|x(\tilde{t})|x\rangle  x} \right] {\rm Tr }_{\rm env} \left[ \langle x|\rho (t)|y\rangle  \right] ,
\end{eqnarray}
where $\rho$ denotes the density operator for the total system including both, trapped particle and heat bath.
Using Eqs. (\ref{eqn:r1}, \ref{eqn:rho2}), we get
\begin{eqnarray}
 \langle x(t+\tilde{t}) x(t)\rangle &=& \int { dx dy dx' \, x\, x'\, J(x',x',x,y,\tilde{t},0)\, \rho_{\rm sys}(x,y,t) } .
 \label{eqn:26}
\end{eqnarray}
Inserting the explicit expressions for $J$ and $\rho$, eqs. (\ref{eq:J}) and (\ref{eqn:1}), we thus obtain
\begin{eqnarray}
 \langle x(t+\tilde{t}) x(t)\rangle  &=& \int{ dX d\xi dX' \,\frac{X'(X+\xi)}{8}
 \rho (X,\xi,t) J ( X',0,X,\xi,\tilde{t},0) } \nonumber \\
&=& \int  dX d\xi dX' \, \frac{X'(X+\xi)}{8} \exp\left[ - \left( b(t) X^2 + a(t) \xi^2 +i c(t) X \xi \right) \right] 
\nonumber \\
&& \times  M'(t) F^2(\tilde{t}) \exp \left[ i K_1(\tilde{t}) X \xi - i N(\tilde{t}) X' \xi - C(\tilde{t}) \xi^2 \right]  ,
\end{eqnarray}
where $M'(t) = F^2 (t) \sqrt{16 \pi M \omega / z(t)}$ and $a$, $b$ and $c$ are given by eqs. (\ref{eqn:a}, \ref{eqn:b}, \ref{eqn:c}). The Gaussian integral that can be solved: 
\begin{eqnarray}
 \langle x(t+\tilde{t}) x(t)\rangle  =  - \frac{F^2 (\tilde{t}) F^2 (t) \sqrt{M \omega}\,  \pi^2}{\sqrt{z(t) b(t)} |N(\tilde{t})| } 
 \left[ \frac{K_1(\tilde{t}) - c(t) }{2 b(t) N(\tilde{t})} - \frac{i}{N(\tilde{t})} \right] .
\end{eqnarray}
Using the normalization condition for the density operator we finally obtain the autocorrelation function for the position operator:
\begin{eqnarray}
 \langle x(t+\tilde{t}) x(t)\rangle  = - {\sqrt{M \omega}\, e^{\gamma(t - \tilde{t})} \over { 4 \sqrt{b(t) z(t)}}} 
 \frac{\sin{ w \tilde{t}}}{|\sin{ w t}|} 
 \left[ i + \frac{2 c(t) - M w \cot(w \tilde{t})- M \gamma }{ 4 b(t) } \right] ,
 \label{eqn:main}  
\end{eqnarray}
where $ \omega $ is the natural frequency of the oscillator and $ w $ is given by 
\begin{equation} 
w^2 = \omega^2 - \gamma \left( \frac{4 \Omega }{\pi} + \gamma \right) .
\end{equation}
and  and $z$, $b$ and $c$ are given by eqs. (\ref{eqn:z}, \ref{eqn:b}, \ref{eqn:c}).

\subsection{Position uncertainty} \label{sec:5}

We first consider the equal-time position uncertainty. Substituting $\tilde{t} = 0 $ in eq.  (\ref{eqn:main}), we find 
\begin{eqnarray}
\langle  x^2 (t) \rangle = \frac{z(t)}{8 {N(t)}^2 M \omega } \label{eqn:29}
\end{eqnarray}
It is instructive to check the correctness of this result at $t=0$ and in the limit $t\to\infty$.

For $t=0$, $\langle x^2 (t)\rangle$ given by eq. (\ref{eqn:29}) has to reduce to the position uncertainty for the chosen initial state of the system oscillator. The position uncertainty of a free harmonic oscillator in its ground state is given by 
\begin{eqnarray}
 \langle  x^2 \rangle_0  = \frac{1}{2 M \omega}  \label{eqn:30}
\end{eqnarray}

On the other hand, inserting $t=0$ in (\ref{eqn:29}), we find:
\begin{eqnarray}
\langle  x^2 (0) \rangle  &=& \frac{1}{2 M \omega} \frac{z(0)}{4 N^2 (0)} \nonumber \\ 
&=& \frac{1}{2 M \omega} \left.  \frac{ 4 {K_1(t)}^2 + M^2 \omega^2 + 4 M \omega C(t) } {{M^2 \omega^2 \over \sin^2{\omega t})} } \right|_{t=0} \nonumber 
\end{eqnarray}
Using $  C(0) = 0 , K_1(t) \to \frac{1}{2} (M \omega \cot{\omega t}+ M \gamma)$ for $t\to 0$ above, we get 
\begin{eqnarray}
\langle  x^2 (0) \rangle   &=&  \frac{1}{2 M \omega} \left. \frac{ ( M \omega  \cos{\omega t } + M \gamma \sin^2{\omega t}) + M^2 \omega^2 (\sin{\omega t})^2}{M^2 \omega^2}\right|_{t=0}\nonumber \\
&=& \frac{1}{2 M \omega} \left.  \frac{M^2 \omega^2 (\cos^2{\omega t})}{M^2 \omega^2}\right|_{t=0}
=  \frac{1}{2 M \omega} ,
\end{eqnarray}
which agrees with (\ref{eqn:30}).

In the limit $ t \rightarrow \infty $, the position uncertainty of the system oscillator should reduce to that for the harmonic oscillator in thermal equilibrium with its environment. For a simple harmonic oscillator in thermal equilibrium, the position uncertainty is given by
\begin{eqnarray}
 \langle x^2\rangle_{\rm th} 
= \frac{{\rm Tr}(x^2 \rho_{\rm th} )}{{\rm Tr}(\rho_{\rm th})} 
= \frac{\sinh(w / \theta)}{2 M w ( \cosh(w / \theta) - 1 )} 
= \frac{\coth{\left[{w / ( 2 \theta) } \right]}}{2 M \omega}
\label{eqn:31}
\end{eqnarray}
On the other hand, taking the long time limit of (\ref{eqn:29}), we obtain:
\begin{equation}
\lim_{t\to\infty}  \langle  x^2 (t) \rangle =  \frac{4 K_1(t)^2+ 4 M \omega C + M^2 \omega^2 }{8 M \omega M^2 w^2} \exp{(-2 \gamma t)} 4 {(\sin{w t})}^2
\end{equation}
Using the explicit forms of $ K_1 (t) $ and $ C(t) $ the leading term is:
\begin{eqnarray}
\langle x^2(t) \rangle & \to & 
 \frac{2}{M \pi} \int_0^ \Omega v dv\, \coth\frac{v}{2 \theta} \left[ \frac{1}{ (v+w)^2 + \gamma^2} \right] \left[ \frac{\gamma}{ (v-w)^2 + \gamma^2} \right]
\end{eqnarray}
In the limit $\gamma \rightarrow 0$ the last factor becomes a delta function. Further taking $\Omega \rightarrow \infty$ we obtain
\begin{eqnarray}
\langle x^2(t) \rangle 
& \to &  \frac{2}{M \pi} \int_0^ \Omega v dv\, \coth\frac{v}{2 \theta} \left[ \frac{1}{ (v+w)^2 + \gamma^2} \right]  
\pi \delta(v-w) \nonumber \\
&=& \frac{\coth{\left[{w / ( 2 \theta) } \right]}}{2 M w} ,
\label{eqn:inf0}
\end{eqnarray}
which agrees with (\ref{eqn:31}).

\subsection{Unequal-time correlator}

We now turn to the unequal time correlator $ \langle x(t') x(t) \rangle$ of the position of the system oscillator. Interpreting the quantum oscillator as a $(0+1)$-dimensional quantum field theory, this quantity is the analog of the two-point function of the field operator.

\subsubsection{ Long-time limit }

We find the two point correlation function $ \langle x(\tilde{t}) x(0) \rangle_{\rm th} $ for a harmonic oscillator in thermal equilibrium with a reservoir at a temperature $ \theta $:
\begin{eqnarray}
\langle x(\tilde{t}) x(0) \rangle_{\rm th} 
&=& {\rm Tr} \left[ \rho_{\rm th} U^{\dagger}(\tilde{t}) x(0) U(\tilde{t}) x(0) \right] \nonumber \\
&& = \int dx dy dz\, x\, z\, \rho_{\rm th} (x,y) \langle y|U^{\dagger}(\tilde{t})|z\rangle  
\langle z | U(\tilde{t})|x\rangle  
\label{eqn:t0}
\end{eqnarray}
The density matrix of a thermal harmonic oscillator in position space is:\cite{3}
\begin{eqnarray}
 \rho_{\rm th} (x,y) &=& \langle x | \exp( -\beta H ) | y \rangle \nonumber \\
&=& \sqrt{\frac{M w}{\pi} \tanh\frac{w}{2 \theta}}\,
  \exp{\left[ -\frac{M w } { 2 \sinh(w / \theta)} \left( (x^2 + y^2 ) \cosh\frac{w}{ \theta} - 2 x y \right)\right]} .
\end{eqnarray}
The time evolution operator has the position matrix elements:
\begin{eqnarray}
 \langle z |U(\tilde{t})| x \rangle &=& \sqrt{\frac{M w}{2 \pi i \sin{w \tilde{t}}}}\, 
 \exp{\left[ -\frac{ i M w } { 2 \sin{w \tilde{t} }} \left( (z^2 + x^2 ) \cos{w \tilde{t}} - 2 x z \right) \right]} 
\end{eqnarray}
Using these in (\ref{eqn:t0}) and solving the Gaussian integral, we get 
\begin{eqnarray}
  \langle x(t+\tilde{t}) x(t) \rangle_{\rm th} &=&
  \langle x(\tilde{t}) x(0) \rangle_{\rm th} 
  = \frac{1}{2 M w } \left[  \cos{(w \tilde{t} )  } \coth\left( {w \over 2 \theta } \right) - i  \sin{ w \tilde{t}}\right]
\end{eqnarray} 
where, as before,  $w$ is the effective frequency of oscillation of the damped system oscillator.

We now compare the long-time limit (\ref{eqn:main}) of the C-L system oscillator with the thermal oscillator. For the parameter choices
\begin{equation}
\theta = 10 \mbox{~MeV} , 
\gamma = 0.2  \mbox{~MeV}  , 
\Omega = 100 \mbox{~MeV} , 
M = 130 \mbox{~MeV} ,
\end{equation}
 we find
\begin{eqnarray}
  \langle x(t+\tilde{t}) x(t) \rangle_{\rm th} &=& 0.001736  {\mbox{~(MeV)}}^{-2} \nonumber  \\
\langle x(t+\tilde{t}) x(t)\rangle_{\rm sys} &=& 0.001734  {\mbox{~(MeV)}}^{-2} 
\end{eqnarray}
for $ \tilde{t} = -i / 5  {~\mbox{(MeV)}}^{-1} $ and $ t = 20  {~\mbox{(MeV)}}^{-1} $.

Thus we see that the autocorrelation function of the position operator at different times shows the expected behavior.

\section{Imaginary time correlation function}

\subsection{Periodicity of the thermal correlator}

It is well known\cite{2} that the two-point function of any Heisenberg operator  $O$, $\langle O(\tilde{t}))O(0)\rangle$ for a thermal system at a temperature $ \theta $ is periodic in imaginary time $\tilde{t}=-i\tau$ with the period $\beta = 1/\theta$. Consider the thermal average of the unequal-time correlation function of any operator $O$:
\begin{eqnarray}
\langle O(\tilde{t}))O(0)\rangle_{\rm th}  
&=& {\rm Tr} \left[ \rho_{\rm th} O(\tilde{t}))O(0) \right] \nonumber \\
&=& {\rm Tr} \left[ \exp{(-\beta H)} \exp{(i H \tilde{t})} O(0) \exp{(-i H \tilde{t})} O(0) \right] 
\end{eqnarray}
For $\tilde{t} = -i \beta$ one thus obtains, making use of the invariance under cyclical permutation of the operators under the trace:
\begin{eqnarray}
 \langle O(-i\beta))O(0)\rangle_{\rm th} 
&=& {\rm Tr} \left[ \exp{(-\beta H)} \exp{(\beta H)} O(0) \exp{(- \beta H)} O(0) \right]  \nonumber \\
&=& {\rm Tr} \left[ O(0)) \exp{(- \beta H)} O(0) \right]  \nonumber \\
&=& {\rm Tr} \left[ \exp{(- \beta H)} O(0))O(0) \right]  
= \langle O(0) O(0)\rangle_{\rm th} .
\end{eqnarray}
We note in passing that this property only implies periodicity of the correlation function on the toroidally compactified interval $\tau \in [0,\beta]$, not on the entire imaginary time axis. For example:
\begin{eqnarray}
 \langle O(-2i\beta))O(0)\rangle_{\rm th} 
&=& {\rm Tr} \left[ \exp{(-\beta H)} \exp{(2\beta H)} O(0) \exp{(- 2\beta H)} O(0) \right]  \nonumber \\
&=& {\rm Tr} \left[ \exp{(\beta H)} O(0)) \exp{(- 2\beta H)} O(0) \right]  \nonumber \\
&=& {\rm Tr} \left[ \exp{(- 2\beta H)} O(0)) \exp{(\beta H)}O(0) \right]   \nonumber \\
&=& {\rm Tr} \left[ \exp{(- \beta H)} \exp{(- \beta H)} O(0)) \exp{(\beta H)}O(0) \right]   \nonumber \\
&=& \langle O(i\beta) O(0)\rangle_{\rm th}  \nonumber \\
&\neq &  \langle O(-i\beta))O(0)\rangle_{\rm th} .
\end{eqnarray}
In fact, one easily checks that the ``periodicity'' property generalizes to the statement that the thermal imaginary time correlation function is symmetric under reflection at the point $\tilde{t}=-i\beta/2$, rather than to periodicity along the full imaginary time axis:
\begin{eqnarray}
 \langle O(-i\beta/2+\tilde{t})) && \!\!\! O(0)\rangle_{\rm th} \nonumber \\
&&= {\rm Tr} \left[ \exp{(- \beta H)} \exp{(\beta H/2+iH\tilde{t})} O(0)) \exp{(-\beta H/2-iH\tilde{t})} O(0) \right] 
\nonumber \\
&&= {\rm Tr} \left[ \exp{(-\beta H/2+iH\tilde{t})} O(0)) \exp{(-\beta H/2-iH\tilde{t})} O(0) \right]  \nonumber \\
&&= {\rm Tr} \left[ \exp{(-\beta H/2-iH\tilde{t})} O(0)) \exp{(-\beta H/2+iH\tilde{t})} O(0) \right]  \nonumber \\
&&= {\rm Tr} \left[ \exp{(- \beta H)} \exp{(\beta H/2-iH\tilde{t})} O(0)) \exp{(-\beta H/2+iH\tilde{t})} O(0) \right]  
\nonumber \\
&&= \langle O(-i\beta/2-\tilde{t}) O(0)\rangle_{\rm th}  .
\end{eqnarray}

\subsection{Periodicity of the correlator in the long-time limit}

If the system oscillator thermalizes in the limit $t\to\infty$, the periodicity in imaginary time $\tilde{t}$ must also be valid for the two-time correlation function $\langle x(t+\tilde{t}) x(t)\rangle$ for the system for large values of $t$:
\begin{equation}
 \lim_{t\to\infty} \langle x(t-i\beta) x(t)\rangle = \lim_{t\to\infty} \langle x(t) x(t)\rangle .
\end{equation}

We can easily check the approximate periodicity numerically for a large real value of $t$:
For example, choosing
$$ 
\theta = 10 \mbox{~MeV}, \gamma = 0.2  \mbox{~MeV} , 
\Omega = 100 \mbox{~MeV},  M = 130 \mbox{~MeV} , 
t= 200 \mbox{~(MeV)}^{-1} 
$$
we find:
\begin{eqnarray}
\langle x(t) x(t)\rangle  &=&  0.008323  \mbox{~(MeV)}^{-2} , \nonumber \\
\langle x \left( t-i/\theta \right) x(t)\rangle &=& 0.008323  \mbox {~(MeV)}^{-2}  .
\end{eqnarray}

An analytical proof can be obtained as follows. Substituting $ \tilde t= -i\beta$ in (\ref{eqn:main}) and using the fact that $ c(t) \rightarrow 0$  as $t \rightarrow \infty $, we find
\begin{eqnarray}
  \langle x \left( t-i\beta \right) x(t)\rangle 
  =  - \frac{\sqrt{M \omega}\, e^{\gamma(t + i\beta})}{ 4 \sqrt{b(t) z(t)}} 
 \frac{\sin{ \left( { -i w \beta}\right) }}{|\sin{ w t}|}  
 \left[ i - \frac{ M w \cot\left( -i w \beta \right)+ M \gamma }{ 4 b(t) } \right] .
 \end{eqnarray}
Substituting $b(t) z(t) = M N^2(t) \omega $ in (\ref{eqn:b}) we get: 
\begin{eqnarray}
  \langle x \left( t-i\beta \right) x(t)\rangle 
  =  \frac{\sqrt{M \omega}\, e^{\gamma t} e^{i \gamma \beta}} { 4 {\sqrt{M \omega}}|N|} 
 \frac{\sin{ \left( { -i w \beta}\right) }}{|\sin{ w t}|}  
 \left[- i + \frac{ M w \cot\left( { -i w \beta}\right)+ M \gamma }{ 4 b(t) } \right] .
\end{eqnarray}
Further using the expression $ N(t)= {M w e^{\gamma t} \over 2 | sin(wt)|} $ and simplifying, we get:
\begin{eqnarray}
  \langle x \left( t-i\beta \right) x(t)\rangle =  \frac{e^{i \gamma \beta}} { 2 M w}
\left[ - \sinh{ \left( { w \beta}\right) } + \frac{ M w \cosh\left( { w \beta}\right) - i M \gamma \sinh{ \left( { w \beta}\right) }}
  { 4 b(t) } \right] 
  \label{eqn:prev1}
\end{eqnarray}
Now, 
\begin{equation}
 \frac{1}{4 b(t)} = \frac{z}{4 N^2 M \omega} = 2 \langle x^2(t) \rangle .
\end{equation}
Using the long-time limit (\ref{eqn:inf0}) for the position variance and taking the limit $\gamma \to 0$,  (\ref{eqn:prev1}) becomes:
\begin{eqnarray}
  \langle x \left( t-i\beta \right) x(t)\rangle
&=& { 1 \over { 2 M w}} \left[\coth{\left( w \beta/2 \right)}  \cosh\left( w \beta \right)-\sinh\left( w \beta \right) \right]\nonumber \\
&=& \frac{ \coth \left( w \beta/2 \right)} { 2 M w} \left[\cosh ( w \beta )-\tanh ( w \beta/2 ) 
\sinh ( w \beta ) \right] \nonumber \\
&=& \frac{ \coth ( w \beta/2 ) } { 2 M w} . 
\label{eqn:p2}
\end{eqnarray}
In the last step we used the relations $\cosh(x) = 1 + 2\sinh^2(x/2)$ and $\sinh(x) = 2 \sinh(x/2)\cosh(x/2)$.
This coincides with the result (\ref{eqn:inf0}) for $ \langle x^2(t) \rangle $ and thus confirms the periodicity of the two-time correlation function in the long-time limit.

\subsection{Correlation function in imaginary time}

\begin{figure}[htb]
\centering
\includegraphics[scale=0.75]{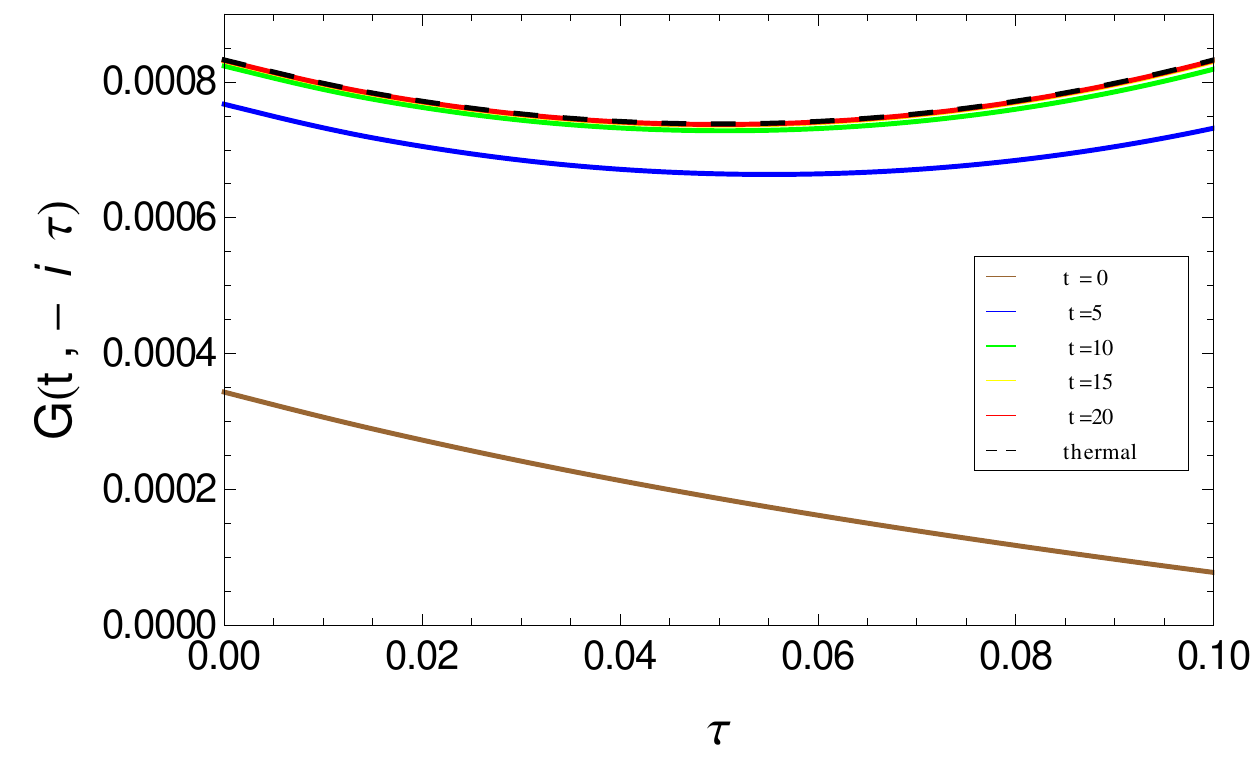}
\caption{Periodicity of the two-point function $G(t,-i\tau)$ on the interval $0 \leq \tau \leq \beta$ for $\theta = 10 \mbox{~MeV}$. The brown, blue, green, and red lines correspond to times $t=0,5,10,20 \mbox{~(MeV)}^{-1}$, respectively. For times of the order of the thermalization time, the two-point function approaches periodicity. The thermal correlation function is shown as a black dashed line.}
\label{fig:i12}
\end{figure}

\begin{figure}[htb]
\centering
\includegraphics[scale=0.75]{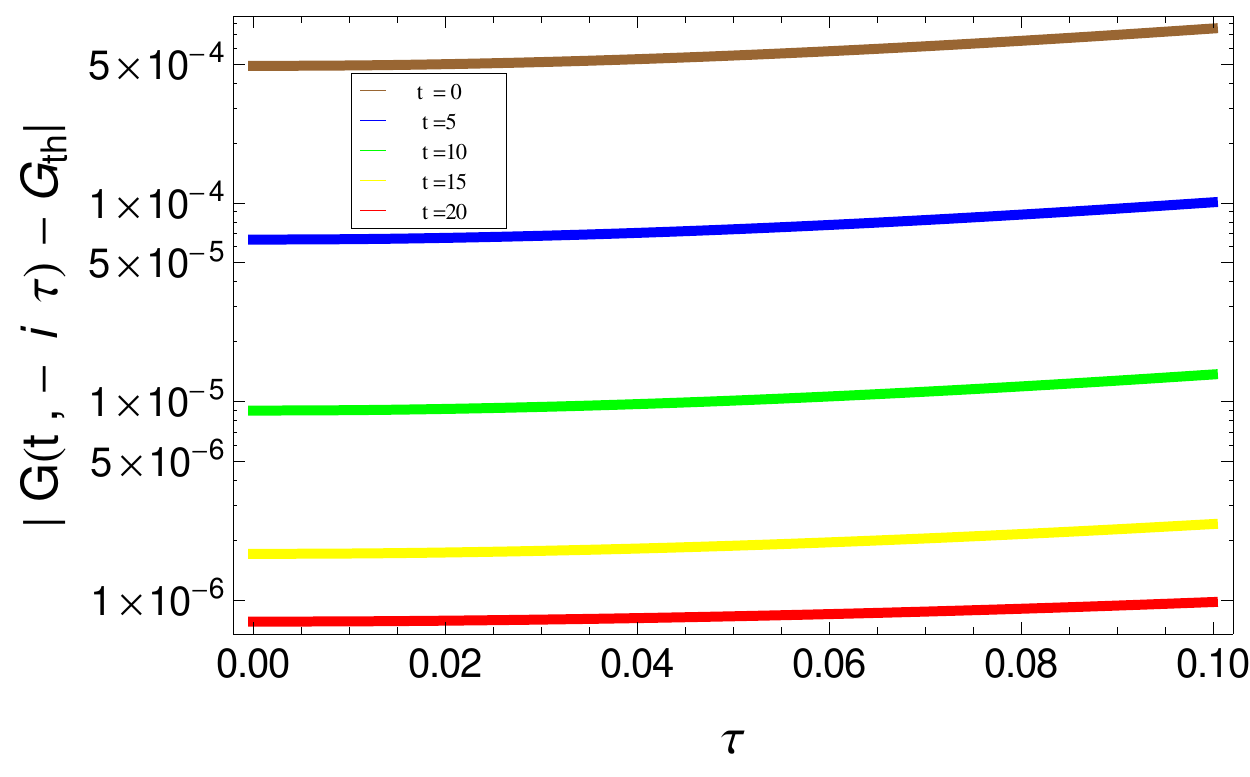}
\caption{Deviation of the imaginary time two-point  function for the C-L model from the thermal correlator in a semilogarithmic plot. The various curves are for different values of the start time $t$ of the correlator.}
\label{fig:i4}
\end{figure}

The Figures \ref{fig:i12} -- \ref{fig:i4} below explore the behavior of the two-time correlation function 
\begin{equation}
G(t,\tilde{t})= \langle x(t+\tilde{t})x(t)\rangle
\end{equation} 
in the Caldeira-Leggett model for imaginary values of the time difference $\tilde{t}=-i\tau$.  The differently colored lines represent the system oscillator at various times $t$. The black dashed line always represents the thermal system.

The approach of the imaginary-time correlation function of the C-L model towards the thermal correlator is obvious from these figures. The periodicity of the late-time two-point function on the compactified interval $0 \leq \tau \leq \beta$ is clearly visible in Fig \ref{fig:i12}. The smooth approach to the thermal correlator is shown in Fig. \ref{fig:i4}. 

\begin{figure}
\centering
\includegraphics[scale=0.75]{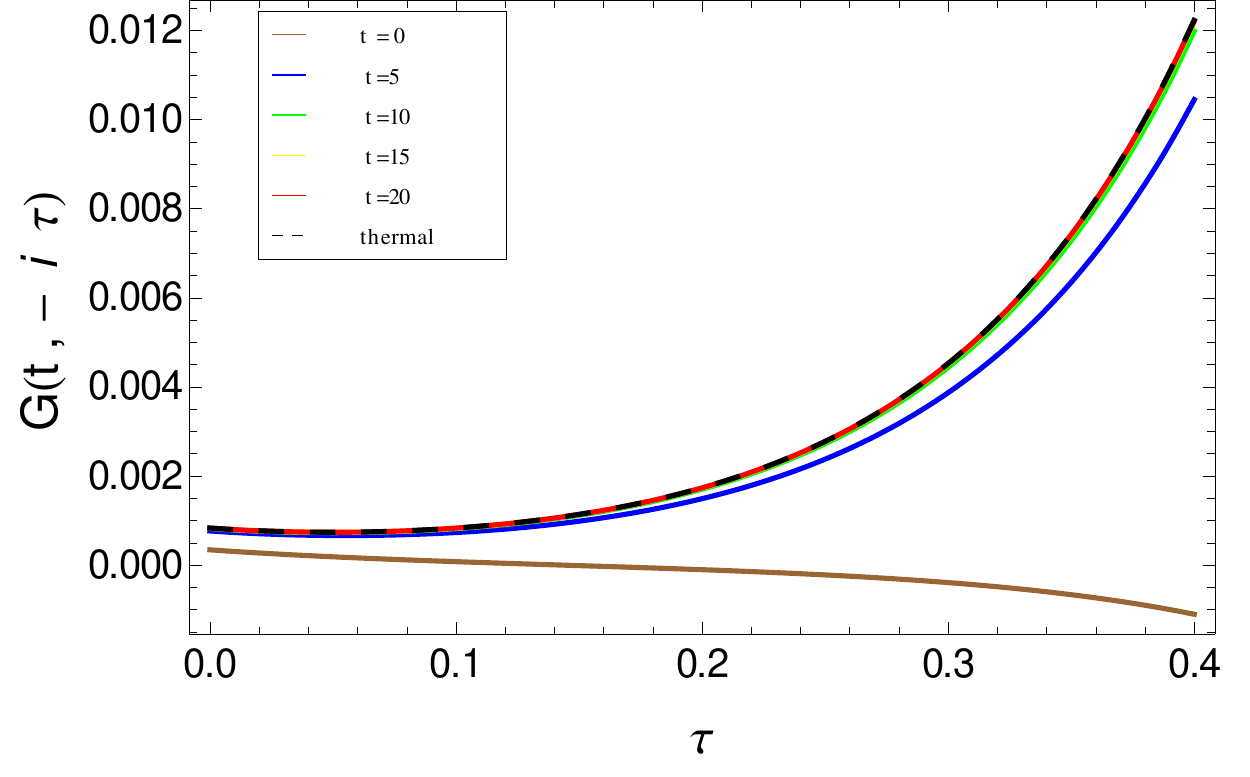}
\caption{Comparison of the two-point function $G(t,-i\tau)$ with the thermal equilibrium correlator on the extended imaginary time interval $0 \leq \tau \leq 0.4$. Near the thermalization time, the two-point function for the system merges with that for a thermal system. The absence of periodicity over the larger time interval is clearly seen.}
\label{fig:i2}
\end{figure}

Figure \ref{fig:i2}, which shows the imaginary time correlation function in comparison with the thermal correlator over a larger imaginary time domain, illustrates the absence of periodicity of the correlation function beyond the interval $0 \leq \tau \leq \beta$. The exponential growth of the correlation function for $\tau > \beta$ can easily be understood by realizing that the thermal correlator can be written in terms of the eigenstates $|n\rangle$ of the harmonic oscillator as
\begin{equation}
\langle x(-i\tau)x(0)\rangle_{\rm th} = \sum_{m,n} | \langle n | x | m \rangle |^2 \exp[-E_n\beta +(E_n-E_m)\tau] .
\end{equation}
Since the position operator connects states $m = n\pm 1$, this implies that the thermal correlator grows like $\exp(\omega\tau)$ at large imaginary times.

\subsection{Correlation function in real time}

\begin{figure}[htb]
\centering
\includegraphics[scale=0.75]{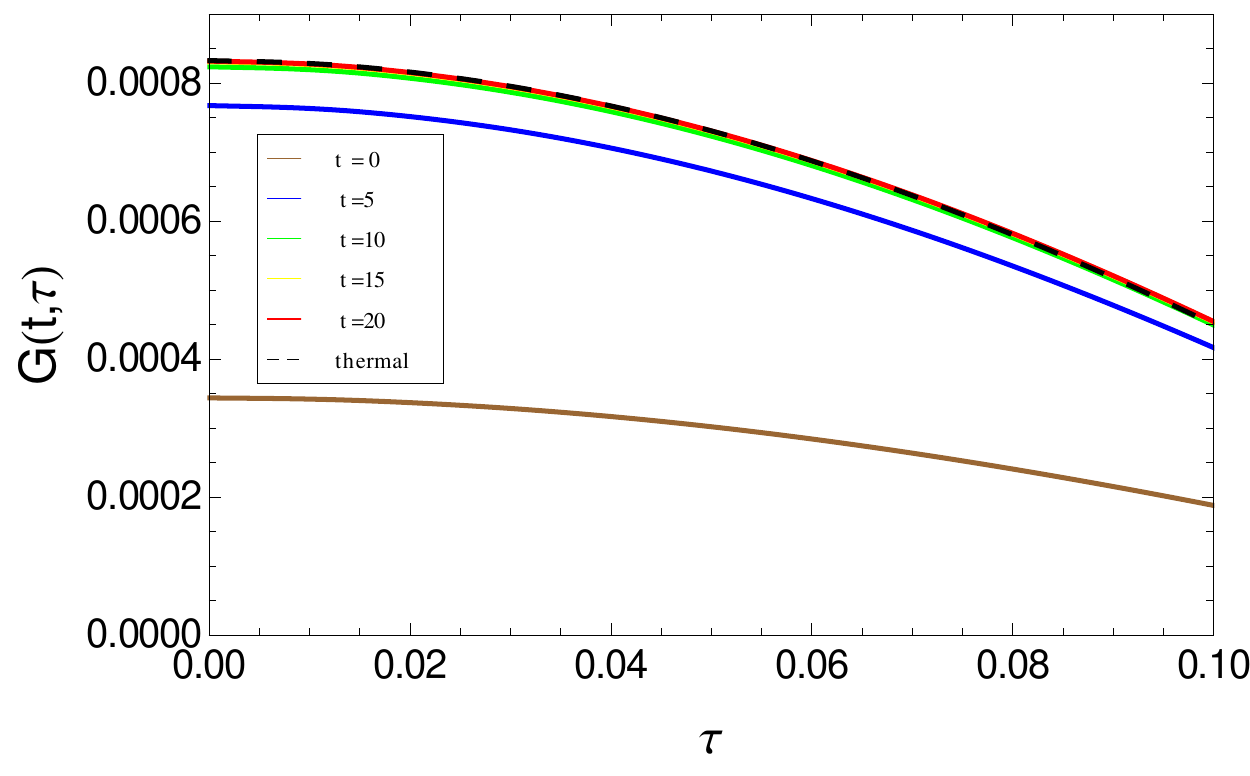}
\caption{Correlation function $G(t,\tau)$ as function of the real time difference $\tau$ in the interval $0 \leq \tau \leq \beta$ for several values of $t$. The brown, blue, green, and red lines correspond to times $t = 0,5,10,20 \mbox{~(Mev)}^{-1}$, respectively.}
\label{fig:r1}
\end{figure}

\begin{figure}[htb]
\centering
\includegraphics[scale=0.75]{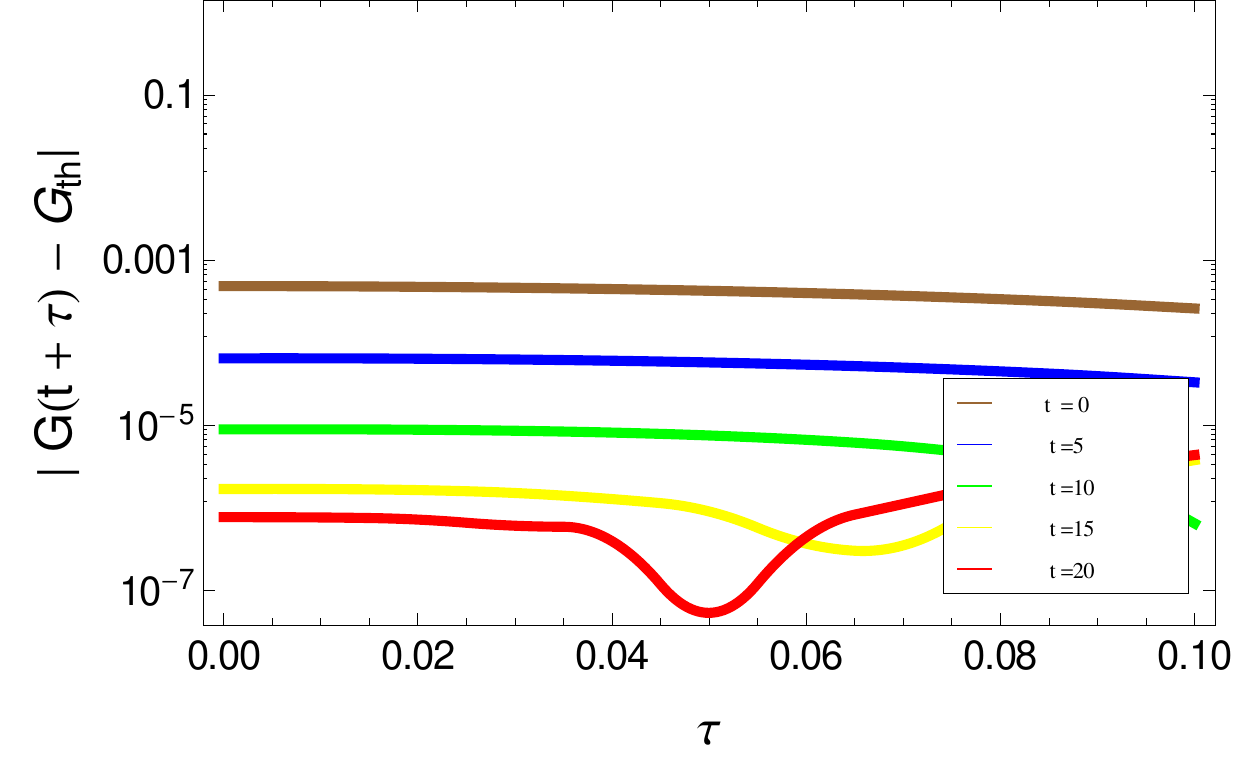}
\caption{Deviation of the correlation function for C-L model from the thermal system in real time $\tau$ in a semilogarithmic plot.}
\label{fig:r4}
\end{figure}

\begin{figure}[htb]
\centering
\includegraphics[scale=0.75]{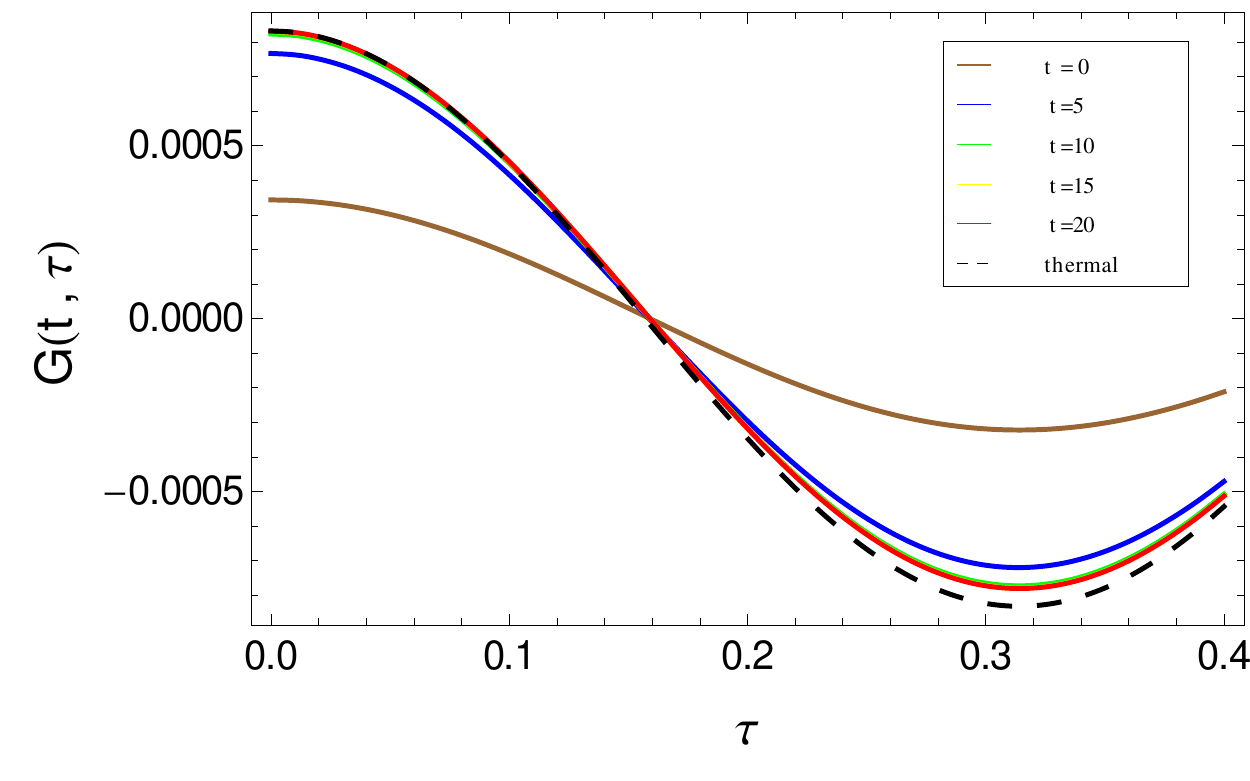}
\caption{Comparison of the two point  function for C-L system with the thermal system in real time . Near the thermalization time, the curve for the system approaches that for a thermal system.}
\label{fig:r2}
\end{figure}

For completeness, we also investigate the dependence of the two-time correlation function on real time difference.
The figures below show the behavior of the two-point  correlation function $ \langle x(t+\tilde{t})x(t)\rangle $ for the Caldeira-Leggett model in real relative time $\tilde{t}$. Figure \ref{fig:r1} shows the correlation function $G(t,\tau)$ as function of the real time difference $\tau$ for several values of $t$. The correlator is seen to approach the thermal value rapidly. Figure \ref{fig:r4} shows the difference between the real-time correlation function of the C-L model and the thermal correlator on a semilogarithmic scale. Finally, Fig. \ref{fig:r2} shows the real-time correlation function over a more extended time interval.

\subsection{ Two point correlation function for complex time}

\begin{figure}[htb]
\centering
\includegraphics[width=0.48\textwidth]{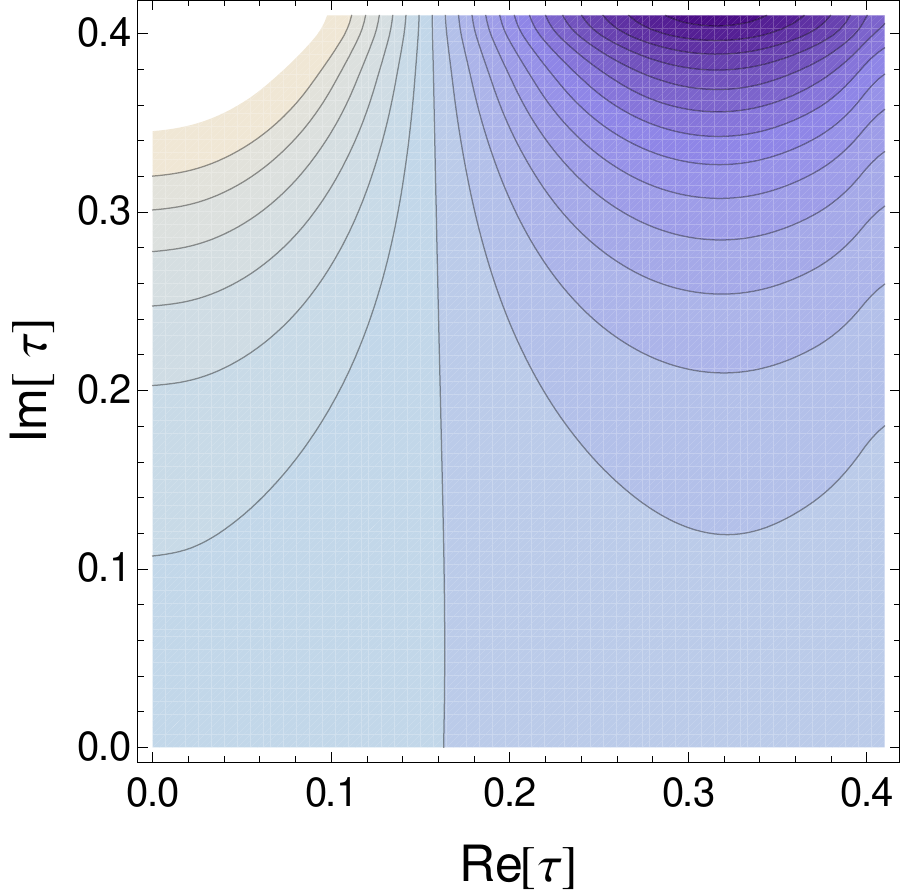}
\hspace{0.01\textwidth}
\includegraphics[width=0.48\textwidth]{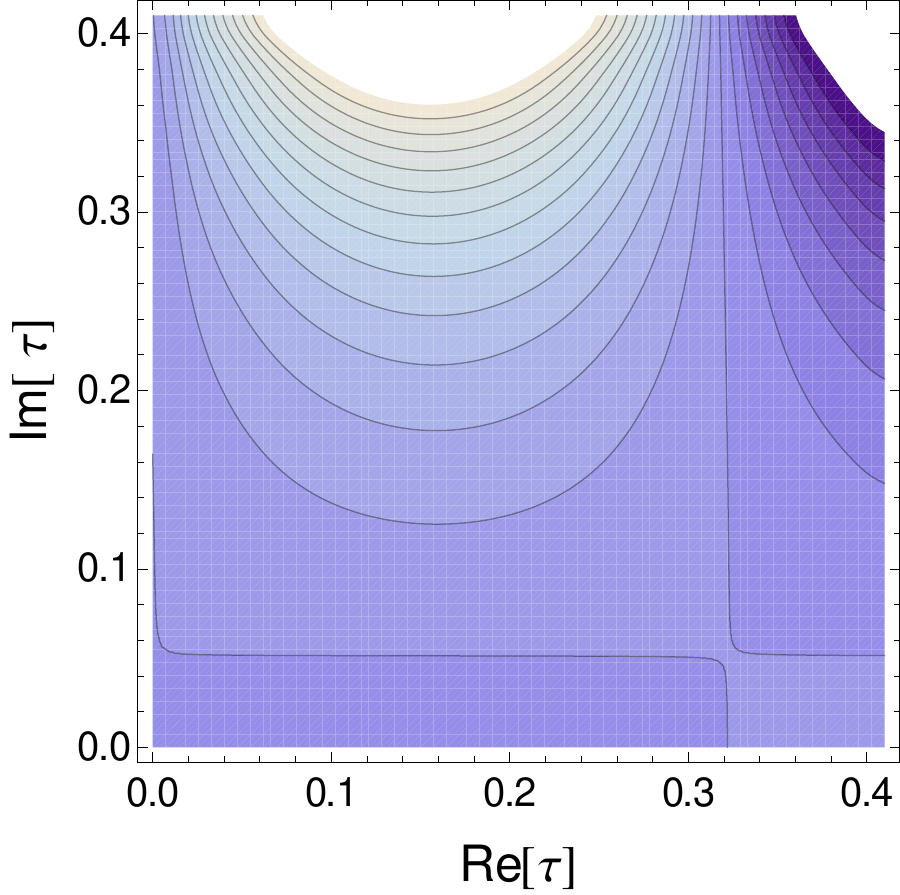}
\caption{ Contour plot of the real part (left) and imaginary part (right) of the correlation function $G(t ,\tau )$ as a function of time $\tau$ in the complex plane. Here, $t = 20 \mbox{~(MeV)}^{-1}$, i.e. the system is thermalized. Lighter shading indicates higher values.}
\label{fig:rx1}
\end{figure}

For completeness, we present contour plots showing the behavior of the real and imaginary parts of the two point correlation function for time $ \tilde{t} $ in the complex plane in Fig.~\ref{fig:rx1}.

\section{ Summary }

We have studied the time evolution of the density matrix of a harmonic oscillator in interaction with a thermal environment, starting in the ground state. Our study makes use of the explicit solution obtained by Caldeira and Leggett when the environment is modeled as a large collection of harmonic oscillators. We confirmed that the  parameters of the solution approach constant values at late times. We also showed that these asymptotic values agree with those for the density operator of the harmonic oscillator in thermal equilibrium. Our comparison with the thermal system shows that the system oscillator of the Caldeira-Legett model thermalizes as expected.

We then computed the two-time correlation function of the position operator for the Caldeira-Leggett model $G(t,\tilde{t}) = \langle  x(t+\tilde{t}) x(t) $, where $\tilde{t}$ denotes the time difference and compared it with the position correlation function for a harmonic oscillator in thermal equilibrium. We studied the following cases in detail:
\begin{enumerate}
\item $\tilde{t}=0$, corresponding to the position uncertainty of the harmonic oscillator representing the ``system''. Here we showed that the uncertainty changes smoothly from that of the harmonic oscillator in its ground state at $t=0$ to that of a thermal oscillator in the long-time limit.
\item Imaginary $\tilde{t}$.  We found that the correlation function $G(t,\tilde{t})$ approaches the thermal correlation function $\langle x(\tilde{t}) x(0) \rangle_{\rm th}$  in the long-time limit $t\to\infty$ and exhibits periodicity in imaginary time $\tilde{t}$, i.~e.\  $\lim_{t\to\infty} [G(t,-i\beta) / G(t,0)] = 1$.
\item Real $\tilde{t}$.  Again, we found that the correlation function $G(t,\tilde{t})$ approaches the thermal correlation function $\langle x(\tilde{t}) x(0) \rangle_{\rm th}$  in the long-time limit $t\to\infty$, but the function does not exhibit periodicity properties along the real $\tilde{t}$-axis.
\end{enumerate}

Our results suggest that the thermalization of a subsystem of a larger quantum system evolving in real time can be explored by calculating two-time correlation functions of the form
\begin{equation}
C(t,\tau) = \langle O(t-i\tau) O(t) \rangle
\label{eq:Corr}
\end{equation}
 of operators defined on the subsystem for real values of $t$ and $\tau$ and then studying the behavior of $C$ as a function of $\tau$ for large values of the real time $t$. If the subsystem thermalizes, $C$ will become periodic in the variable $\tau$, i.~e.\ $C(t,\beta) \to C(t,0)$ for $t\to\infty$. The period $\beta$ is a measure of the temperature $\theta=1/\beta$ to which the subsystem thermalizes.

Our results also raise some interesting questions. For example, if one considers a large system in isolation that ``thermalizes'' under its internal dynamics, will correlation functions of the form (\ref{eq:Corr}) exhibit periodicity in $\tau$ for large values of $t$? Will this only be true for certain operators $O$, which are sensitive only to a small subset of all degree of freedom of the large system, e.~g.\ single-particle observables in a many-body quantum system? For a small quantum system, can the average over many degrees of freedom be replaced with an ergodic average over the observation time $t$? 

\section*{Acknowledgements}

This  work was supported by a research grant from the U. S. Department of Energy (DE-FG02-05ER41367).

\end{document}